# Metal-oxide interface reactions and their effect on integrated resistive/threshold switching in NbO$_x$

Shimul Kanti Nath*, Sanjoy Kumar Nandi, Shuai Li, Robert Glen Elliman
Department of Electronic Materials Engineering, Research School of Physics
The Australian National University, Canberra, ACT 2601, Australia
*Email: Shimul.nath@anu.edu.au

**Abstract:** Reactive metal electrodes (Nb, Ti, Cr, Ta, and Hf) are shown to play an important role in controlling the volatile switching characteristics of metal/Nb$_2$O$_5$/Pt devices. In particular, devices are shown to exhibit stable threshold switching under negative bias but to have a response under positive bias that depends on the choice of metal. Three distinct responses are highlighted: Devices with Nb and Ti top electrodes are shown to exhibit stable threshold switching with symmetric characteristics for both positive and negative polarities; devices with Cr top electrodes are shown to exhibit stable threshold switching but with asymmetric hysteresis windows under positive and negative polarities; and devices with Ta and Hf electrodes are shown to exhibit an integrated threshold-memory (1S1M) response. Based on thermodynamic data and lumped element modelling these effects are attributed to the formation of a metal-oxide interlayer and its response to field-induced oxygen exchange. These results provide important insight into the physical origin of the switching response and pathways for engineering devices with reliable switching characteristics.

## I. INTRODUCTION

Two-terminal metal/oxide/metal (MOM) structures are known to exhibit characteristic resistance changes when subjected to electrical stress and are of interest as active elements in non-volatile memory arrays and adaptive neural networks [1-3]. The resistance changes of interest include both volatile and non-volatile behavior, as well as combinations of these responses [4-6]. Non-volatile resistive switching is observed in a broad range of transition metal oxides and the physical mechanisms responsible are reasonably well understood [7-9]. In contrast, reliable volatile switching is observed in far fewer oxides [10-12] and the understanding is less well advanced. As a consequence, it remains an active area of research with particular attention focused on niobium and vanadium oxides due to their reliable switching characteristics.

Significantly, specific phases of niobium and vanadium oxides (i.e. NbO$_2$, VO$_2$ and V$_2$O$_3$) undergo thermally-induced insulator metal transitions (IMT) [13-15] and there has been particular interest in understanding if these contribute to the volatile switching characteristics [16,17]. Indeed, it is commonly argued that volatile switching in amorphous NbO$_x$ and VO$_x$ films is preceded by crystallization of these phases, and that these are responsible for the observed switching [18,19]. In the case of NbO$_x$ this mechanism has explicitly been invoked to account for an observed 'snap-back' response, in which the device conductivity changes abruptly under current controlled operation [18]. However, other studies have shown that volatile switching in amorphous NbO$_x$ films can be explained by the change in film conductivity caused by local Joule heating [20,21], and that this, combined with the effect of a parallel device resistance, can further explain the observed snap-back response [22,23].

We have recently shown that threshold switching reliability in Nb$_2$O$_5$-based devices can be improved by introducing a reactive metal (e.g. Nb) as one of the electrodes. The devices were found to exhibit reliable threshold switching after an initial electroforming step and had threshold and hold voltages that were independent of film thickness and device area [24,25]. On this basis it was concluded that the active switching volume was localized, likely in the vicinity of the reactive-electrode/oxide interface

[24,25]. As such the switching characteristics are expected to be sensitive to interface reactions and field induced oxygen exchange between the electrode and oxide.

In this paper we explicitly examine the influence of different top electrode (TE) metals on the threshold switching response of amorphous $NbO_x$ films and show that the choice of metal does indeed play an important role in determining the switching characteristics.

## II. EXPERIMENTS

Cross-point test structures were fabricated using a step-by-step lithography process [26]. The bottom electrode consisted of a 5 nm Ti adhesion layer and a 25 nm Pt layer which were deposited by electron-beam evaporation onto a thermally oxidized Si substrate (oxide thickness ~ 200nm) without breaking vacuum. A functional layer of amorphous $NbO_x$ with ~45 nm thickness was subsequently deposited by RF-sputtering from a $Nb_2O_5$ target in an Ar ambient at a working pressure of 4 mTorr, while keeping the total power ~180 W. To complete the MOM structure a 10 nm top electrode (TE) comprising five different metals (Nb, Ti, Cr, Ta and Hf) was deposited on the $NbO_x$ film followed by the deposition of a 25 nm Pt capping layer. In order to compare the role of the reactive electrodes, a set of devices with a 25 nm Pt top electrode (i.e. without the reactive metal layer) was also fabricated.

The stoichiometry of the as-deposited $NbO_x$ film was found to be near $Nb_2O_5$ ($x = 2.6 \pm 0.05$, i.e. slightly over-stoichiometric) as determined by Rutherford backscattering spectrometry (RBS), and to be amorphous as confirmed by grazing incident-angle X-ray diffraction (GIAXRD) and transmission electron microscopy (TEM). Layer thicknesses and compositions were confirmed by transmission electron microscopy (TEM) and Energy Dispersive X-ray (EDX) maps using a JEOL 2100, as shown in Fig. 1 for devices with Nb and Cr TEs. Electrical measurements were performed in atmospheric condition using an Agilent B1500A semiconductor parameter analyser attached to a Signatone probe station.

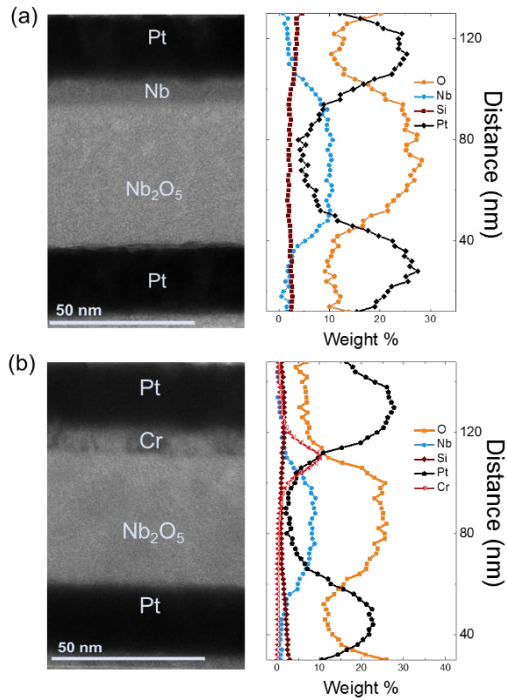

FIG. 1: Cross-sectional transmission micrographs of the device structures: (a) Pt/Nb/$NbO_x$/Pt, and (b) Pt/Cr/$NbO_x$/Pt. Corresponding EDX maps are shown on the right of each image and were obtained from the line scan towards bottom electrode performed from the top electrode or vice versa.

## III.  RESULTS & DISCUSSION

As-fabricated devices were highly insulating (device resistance ~ GΩ) and were subjected to a one-off electroforming process to initiate resistive switching [see the inset of Fig. 2 (a)]. The switching characteristics were initially investigated using current sweeps from 0 to -2.5mA and a region of negative differential resistance (NDR) was evident in the immediate post-forming current-voltage (I-V) sweep for all devices (i.e. for all five top electrode metals). They also exhibited a volatile threshold switching response when subjected to voltage-controlled operation, with typical examples of NDR (solid black line) and threshold switching (dash red line) shown in Fig. 2 (a). Threshold and hold voltages were extracted from the current-controlled NDR characteristics and were found to have mean values of -2.2±0.25 V and -1.6±0.3 V respectively, independent of the top electrode metals as shown in Fig. 2(b).

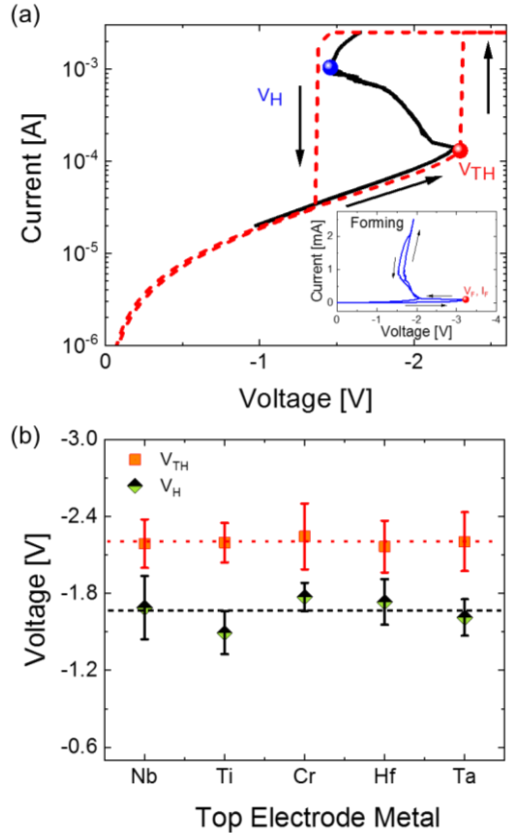

FIG. 2: (a) Current controlled negative differential resistance (NDR) characteristic observed under current-controlled mode (black solid line) and volatile threshold switching (red dashed line) under voltage-controlled operation in a typical device with Nb top electrode; inset shows a typical electroforming step, and (b) threshold- and hold-voltages as a function of different electrodes extracted from the corresponding NDR response (data were taken for 8 devices within each device stack).

Area dependent NDR characteristics were also measured for the Nb TE devices and the threshold and hold voltages were found to be independent of device area (device dimension ranging from 5µm to 20µm). These results are consistent with earlier studies[25,27] and highlight that the electroforming process introduced a conductive filament in the $Nb_2O_5$ film and a localized threshold switching volume was self-assembled at one of the electrode/oxide interfaces[4,28,29].

The switching response was further studied using bidirectional voltage-controlled I-V sweeps from 0 V to -3 V and then from 0 V to +3 V as shown in Fig. 3 and Fig. 4. Stable threshold switching was observed for devices with Nb, Ti and Cr top electrodes (Fig. 3 (a)). In all cases, the sub-threshold I-V characteristics were well represented by a modified Poole-Frenkel mechanism with a trap level of ~0.2 eV. This was determined from temperature dependent I-V measurements using an Arrhenius plot of the low field resistance (measured at 0.5 V), as shown in Fig. 3(b). The obtained trap level (i.e. ~0.2 eV) is consistent with the previous reports [25,30].

On the other hand, devices with Ta and Hf top electrodes exhibited an integrated selector-memory (1S1M) response after a few initial threshold switching cycles as shown in Fig. 4. A detailed explanation of this type of switching can be found in our earlier studies[4,28].

Recent studies have shown that metal electrodes can react with an underlying oxide layer to form a metal-oxide interlayer [5,31]. The extent of the reaction depends on the relative thicknesses of the layers, as well as the choice of electrode metal [31]. Such interaction can alter the switching behaviour, as demonstrated in Figs. 3 and 4.

In order to understand such behaviour we first calculate possible reactions at the top electrode/oxide interface as represented by thermodynamic data for the stable crystalline oxide phases [32-34]. Specifically, we considered the change of Gibbs potential $\Delta G$ calculated using the relation [32]:

$$\Delta G = x_2 \Delta G_2 - x_1 \Delta G_1, \qquad [1]$$

Where, $\Delta G_1$ and $\Delta G_2$ are the standard isobaric potentials of formation of the oxides and $x_1$ and $x_2$ are the number of moles of the oxides which take part in the reaction. From a thermodynamic standpoint, a negative value of $\Delta G$ is associated with a spontaneous reaction and suggests that the metal electrode will react to form an oxide layer and reduce the oxygen content of the $Nb_2O_5$ layer. The calculations summarised in Table-I confirm that this is indeed the case for all five reactive metals used in this study.

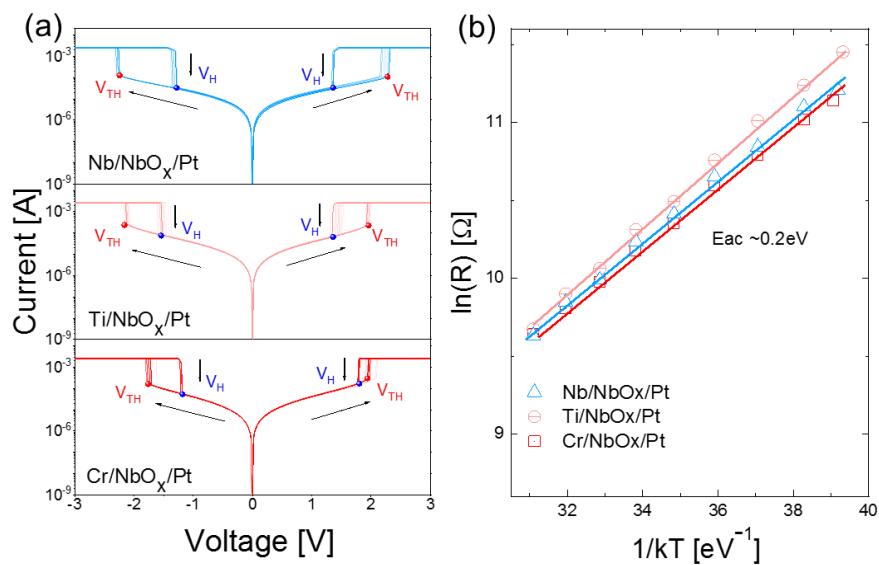

FIG. 3: (a) Voltage-controlled threshold switching in typical devices with Nb, Ti and Cr top electrodes respectively (20 consecutive switching cycles are shown for each case), (b) temperature dependent change of resistance calculated at -0.5 V for electroformed devices with Nb-, Ti- and Cr top electrodes. Note that, while Nb and Ti top electrode devices exhibited symmetric threshold switching, the Cr-top electrode devices exhibited asymmetric threshold switching.

This analysis suggests that an oxide interlayer formed by the reactive metal electrode is likely at the top electrode/$NbO_x$ interface. Moreover, this process is expected to be enhanced during the electroforming step due to local Joule heating [28]. For example, the Gibbs free energy of formation of $HfO_2$ is -381 kJ/mole at room temperature (see Table-I), and the formation of such an interlayer is clearly reasonable at the Hf/NbO$_x$ interface. Since HfO$_2$ is an insulating oxide with a high dielectric constant [35,36], formation of such an interlayer can significantly influence the device characteristics. The resistivity of this interlayer will depend on its thickness and stoichiometry [37,38]. Therefore, any change in these parameters due to local ion migration will also significantly influence the device resistance and its response to the applied electric field.

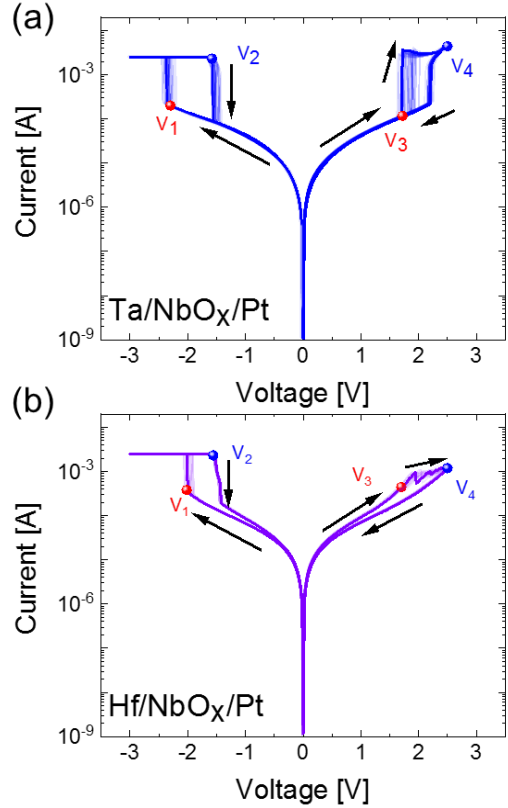

FIG. 4: (a-b) 1S1M behavior observed under voltage-controlled mode in a typical Pt/Ta/NbO$_x$/Pt and a Pt/Hf/NbO$_x$/Pt device respectively (20 consecutive switching cycles for each case).

Based on these arguments the observation of two distinct resistive switching modes (shown in Figs. 3 and 4) can be understood in terms of reactions occurring at the top electrode/NbO$_x$ interface. Since a memory filament is present in both cases, the threshold switching can be explicitly realized by assuming a stable memory filament, and the 1S1M response can be attributed to reconstruction and rupture of the conductive filament by electric field driven oxygen exchange at the top electrode/NbO$_x$ interface.

Table-I: Summary of the MOM stacks and their switching characteristics with different top electrodes.

| MOM Combination | Thermodynamic stability of the top electrode metal/Nb$_2$O$_5$ interface | Stable I-V response |
|---|---|---|
| **Pt − Nb$_2$O$_5$ − Pt** | $5Pt + 2Nb_2O_5 = 5PtO_2 + 4Nb$ ; $\Delta G = +873.253$ kJ/mol | Non-polar Memory/ unstable 1S1M |
| **Nb − Nb$_2$O$_5$ − Pt** | $Nb + 2Nb_2O_5 = 5NbO_2$ ; $\Delta G = -164.265$ kJ/mol | Symmetric TS |
| **Ti − Nb$_2$O$_5$ − Pt** | $5Ti + 2Nb_2O_5 = 5TiO_2 + 4Nb$ ; $\Delta G = -183.062$ kJ/mol | Symmetric TS |
| **Cr − Nb$_2$O$_5$ − Pt** | $2Cr + Nb_2O_5 = Cr_2O_3 + 2NbO$ ; $\Delta G = -38.0455$ kJ/mol | Asymmetric TS |
| **Ta − Nb$_2$O$_5$ − Pt** | $2Ta + Nb_2O_5 = Ta_2O_5 + 2Nb$ ; $\Delta G = -72.5655$ kJ/mol | 1S1M |
| **Hf − Nb$_2$O$_5$ − Pt** | $5Hf + 2Nb_2O_5 = 5HfO_2 + 4Nb$ ; $\Delta G = -381.936$ kJ/mol | 1S1M |

A schematic of the proposed switching elements in a device with Hf top electrode is illustrated in Fig. 5. Here it is assumed that a thin interlayer is formed by a spontaneous reaction between the Hf and $Nb_2O_5$ layers and that this is further enhanced at the vicinity of the electroformed filament due to local Joule heating. The $HfO_x$ layer then acts as a resistive switching layer between the electroformed filament and the top electrode. i.e. When a negative bias is applied to the top electrode Oxygen ions drift from the $HfO_x$ layer into the $NbO_x$ layer to 'set' the $HfO_x$ layer into a conductive state, while under positive bias oxygen-ions are attracted back into the $HfO_x$ layer to 'reset' it into a more resistive state. This combined with the threshold switching response then accounts for the observed 1S1M switching behaviour. Note that, we have assumed that the threshold switching response is dominated by Joule heating of a local volume of material, as previously demonstrated [25].

While this model suggests that 1S1M behaviour might be expected for all reactive electrodes, the resistive switching response will depend on the conductivity of the interlayer, with more resistive, stoichiometry films expected to exhibit the most significant memory switching. The data in Figs. 3 and 4 can then be understood from the fact that Hf and Ta electrodes are likely to lead to highly insulating oxides (e.g. $HfO_2$ and $Ta_2O_5$)[37,38], while Cr, Nb and Ti are more likely to form semiconducting films (e.g. $Cr_2O_3$[39], $NbO_2$[40] and $TiO_2$ [41]). Note that, in addition to the conductivity the resistance of the interlayer will also depend on its thickness.

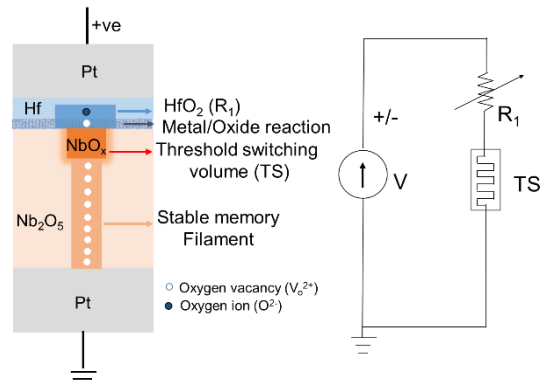

FIG. 5: Schematic of the switching elements in $Pt/Hf/Nb_2O_5/Pt$ device structure after electroforming and a corresponding circuit representation is sketched on the right side. The value of $R_1$ varies under opposite bias conditions due to field driven oxygen transport at the top electrode/oxide interface resulting in the rupture and reconstruction of the memory filament.

Additional insight supporting this assumption can be achieved from the threshold switching behaviour of the Cr-TE devices which have different hysteresis windows ($V_{TH}$-$V_H$) for forward and reverse bias, as shown in Fig. 4(a). This indicates that the interlayer resistance is modified by biasing but not enough to initiate resistive switching. Note that, the threshold switching window (hysteresis) is highly sensitive to any additional series resistance ($R_{Series}$) as depicted in ref. [42] and a small increase in $R_{Series}$ will reduce the window as observed under positive bias in Fig. 4(a). Thus, the corresponding NDR window will also narrow [see Supplementary information].

The switching response of a $Pt/Nb_2O_5/Pt$ structure was also measured for comparison. In this case the devices underwent a few unstable threshold switching or 1S1M cycles before exhibiting unipolar/non-polar switching. The Gibbs free energy of formation of platinum oxide is positive, with a value $\Delta G = +873$ kJ/mol, so no spontaneous reaction is expected at the $Pt/Nb_2O_5$ interface. However, this does not rule out some oxygen exchange during the electroforming process, which may account for the initial transient switching response. Regardless, the clear difference between the response of devices with Pt and reactive-metal electrodes clearly demonstrates the significance of interface reactions in controlling the threshold switching response.

## IV. CONCLUSION

In conclusion, we have examined the role of metal/oxide interface reactions on the threshold switching response of metal/Nb$_2$O$_5$/Pt structures. Reactive metals were found to improve the threshold switching reliability but exhibited characteristic behaviour that was attributed to the nature of the interlayer formed by reaction with the functional Nb$_2$O$_5$ layer.

Specifically, devices with Nb and Ti electrodes exhibited symmetrical threshold switching under positive and negative bias, as expected for low resistance contacts, while those with Cr electrodes exhibited asymmetric switching with a smaller hysteresis window under positive bias than under negative bias, consistent with changes in interlayer resistance induced by field-induced oxygen exchange. In contrast, devices with Hf and Ta electrodes exhibited 1S1M switching, where the 1M response was attributed to resistive switching in a high-resistance HfO$_x$ or TaO$_x$ interlayer, respectively.

Significantly, Pt/Nb$_2$O$_5$/Pt devices exhibited unipolar switching rather than threshold switching, clearly demonstrating the role of the reactive electrodes in mediating the threshold switching response. We further note that the extent of metal/oxide reactions depends on the relative thicknesses of the corresponding material layers thereby providing an additional means of engineering the switching characteristics. These results provide additional insight into the physical origin of the switching response and pathways for engineering devices with reliable switching characteristics.

## ACKNOWLEDGEMENTS

This work was partly funded by the Australian Research Council (ARC) and Varian Semiconductor Equipment/ Applied Materials through an ARC Linkage Project Grant: LP150100693. We would like to acknowledge access to NCRIS facilities at the ACT node of the Australian National Fabrication Facility (ANFF) and the Australian Facility for Advanced ion-implantation Research (AFAiiR). The authors also acknowledge the facilities, and the scientific and technical assistance of the Australian Microscopy & Microanalysis Research Facility at the Centre of Advanced Microscopy, The Australian National University. The authors also thank Dr Tom Ratcliff for comments and feedback on the manuscript

# Supplementary information

**SPICE Models for archetype threshold switching memristor with Poole-Frenkel model**

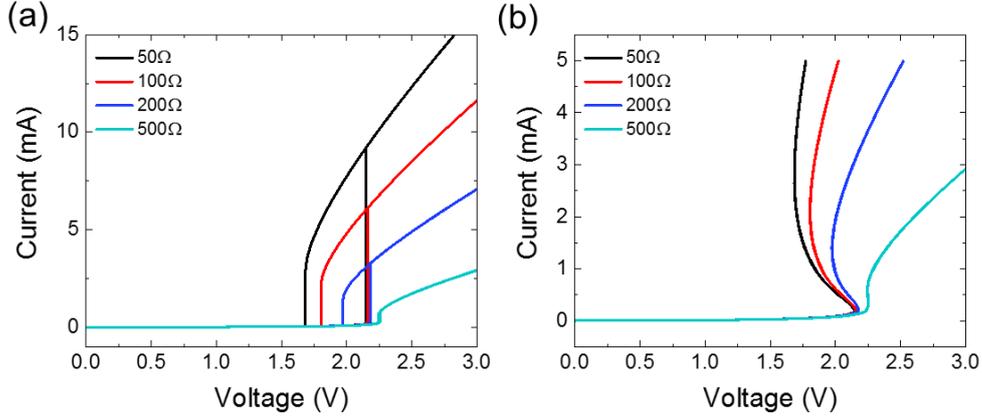

**Figure S1.** Effect of series resistance on the (a) threshold switching hysterestis and, (b) negative differential resistance window. The device was modelled with a threshold switch based on Poole-Frenkel conduction as described in Ref. [22].

We considered an archetype threshold switch (memristor) in our model. The electrical conductivity of the core region was assumed to be Poole-Frenkel type such that the device resistance is given by[20]:

$$R_m = R_0 e^{\frac{1}{k_B T}(E_a - q\sqrt{\frac{qE}{\pi \varepsilon_0 \varepsilon_r}})} \quad .. \quad .. \quad .. \quad (3)$$

where $k_B$ is the Boltzmann constant, $E_a$ is the activation energy, $\varepsilon_0$ is the vacuum permittivity, and $\varepsilon_r$ is the relative permittivity of the threshold switching volume. $T_m$ and $T_{amb}$ denote the temperature of the electrically active region and the ambient environment, $R_0$ is the resistance pre-factor of the active region at $T = T_{amb}$.

The dynamic behaviour of the memristor is defined by Newton's law of cooling:

$$\frac{dT_m}{dt} = \frac{I_m^2 R_m}{C_{th}} - \frac{\Delta T}{R_{th} C_{th}} \quad .. \quad .. \quad .. \quad (4)$$

where $R_{th}$ and $C_{th}$ are the thermal resistance and the thermal capacitance of the device, and $\Delta T$

Details of the model parameters are given in the Table-S1.

**Table S1.** Memristor parameters used in simulation for n the Poole-Frenkel model.

| Model Parameters (Unit) | Symbol | Threshold Switch |
|---|---|---|
| Thermal capacitance (J·K$^{-1}$) | $C_{th}$ | $1\times10^{-15}$ |
| Resistance prefactor (Ω) | $R_0$ | 105 |
| Thermal resistance (K·W$^{-1}$) | $R_{th}$ | $2\times10^5$ |
| Ambient temperature (K) | $T_{amb}$ | 298 |
| Metallic state resistance (Ω) | $R_v$ | 0 |
| Activation Energy (eV) | $E_a$ | 0.23 |

| Boltzmann constant (J·K$^{-1}$) | $k_B$ | $1.38\times10^{-23}$ |
|---|---|---|
| Elementary charge (C) | e | $1.6\times10^{-19}$ |
| Vacuum permittivity (F·m$^{-1}$) | $\varepsilon_0$ | $8.85\times10^{-12}$ |
| Relative permittivity of the threshold switching volume and surrounding NbOx film | $\varepsilon_r$ | 45 |
| Film thickness (nm) | | 45 |
| Series resistance (Ω) | | 50-500 |